\documentclass{article}
\usepackage{spconf,graphicx,amsmath, url}

\usepackage{graphicx} % Required for inserting images
\usepackage{booktabs}
\usepackage{amsmath, amssymb} % for in R 
\usepackage{siunitx}
\usepackage{balance}
\usepackage[sort]{cite}
\usepackage{multirow}
\usepackage[font=small,skip=6pt]{caption}
\usepackage{tikz}
\usepackage{acronym}
\acrodef{DNN}{deep neural network}
\acrodef{TCN}{temporal convolution network}
\acrodef{DPRNN}{dual-path recurrent neural network}
\acrodef{LSTM}{long short-term memory}
\acrodef{TI}{time-invariant}
\acrodef{TV}{time-varying}
\acrodef{TSE}{target speaker extraction}
\acrodef{SIR}{signal-to-interference ratio}
\acrodef{SI-SDR}{scale-invariant source-to-distortion ratio}
\acrodef{ASR}{automatic speech recognition}
\acrodef{SA}{simple attention}
\acrodef{NA}{normalized attention}
\acrodef{gLN}{global layer normalization}
\acrodef{bn}{batch normalization}
\acrodef{cLN}{cumulative layer normalization}
\acrodef{LN}{layer normalization}
\acrodef{SI-SDRi}{scale-invariant source-to-distortion ratio improvement}
\acrodef{VoTSE}{video-only TSE}
\acrodef{AoTSE}{audio-only TSE}
\acrodef{MTSE}{multi-modal TSE}
\acrodef{MTT}{multi-task training}
\acrodef{MDT}{modality dropout training}
\acrodef{SD}{standard deviation}

\title{Training Strategies for Modality Dropout Resilient \\ Multi-Modal Target Speaker Extraction}

\name{Srikanth Korse, Mohamed Elminshawi, Emanu\"{e}l A.~P.~Habets, Srikanth Raj Chetupalli}
\address{International Audio Laboratories Erlangen$^\dag$\thanks{$^\dag$A joint institution of the Friedrich-Alexander-Universit\"{a}t Erlangen-N\"{u}rnberg (FAU) and Fraunhofer IIS, Germany.}, Am Wolfsmantel 33, 91058 Erlangen, Germany}

\begin{document}
\ninept
\maketitle
\sloppy

\begin{abstract}
The primary goal of \ac{MTSE} is to extract a target speaker from a speech mixture using complementary information from different modalities, such as audio enrolment and visual feeds corresponding to the target speaker. \ac{MTSE} systems are expected to perform well even when one of the modalities is unavailable. In practice, the systems often suffer from modality dominance, where one of the modalities outweighs the others, thereby limiting robustness. 
%\Ac{MTSE} systems suffer from modality dominance, where one of the modalities outweighs the others, and they are often expected to perform well even when a modality is unavailable. 
Our study investigates training strategies and the effect of architectural choices, particularly the normalization layers, in yielding a robust \ac{MTSE} system in both non-causal and causal configurations. 
In particular, we propose the use of \ac{MDT} as a superior strategy to standard and \ac{MTT} strategies.
%We also investigate a simple training strategy, called \ac{MDT}, to improve the robustness of~\ac{MTSE} to unavailable auxiliary information. 
Experiments conducted on two-speaker mixtures from the LRS3 dataset show the \ac{MDT} strategy to be effective irrespective of the employed normalization layer. In contrast, the models trained with the standard and~\ac{MTT} strategies are susceptible to modality dominance, and their performance depends on the chosen normalization layer. Additionally, we demonstrate that the system trained with \ac{MDT} strategy is robust to using extracted speech as the enrolment signal, highlighting its potential applicability in scenarios where the target speaker is not enrolled.
\end{abstract}


\acresetall

\section{Introduction}\label{sec:intro}
\vspace{-2.5mm}
The ability of humans to listen to a single speaker in an environment with interfering acoustic sources is commonly referred to as the \textit{cocktail-party effect}~\cite{bronkhorst2015cocktail}. Humans use auxiliary information, such as spatial and visual cues as well as speaker familiarity, to selectively attend to auditory stimuli~\cite{bronkhorst2015cocktail}. However, it is challenging for human-machine interfaces to perform well in these scenarios. 

\Ac{TSE} using \acp{DNN} enables a machine to mimic this complicated task of recovering the speech of a desired speaker (target speaker) from a multi-speaker audio mixture with the help of auxiliary information about the target speaker. %When we refer to modalities, hereafter, we refer to the modalities of the auxiliary information. 
Based on the number of modalities from which the auxiliary information is derived,~\ac{TSE} systems can be classified into either uni-modal or multi-modal systems.
One example of an uni-modal \ac{TSE} system is an \ac{AoTSE} system~\cite{delcroix2018single, Delcroix2020, zhang2020x, xu2020spex}, which utilizes a speech snippet from the target speaker, often referred to as the enrolment signal. Another example is a \Ac{VoTSE} system~\cite{wu2019time, Pan_MUSE_2021_ICASSP, Pan22_USEV_TASLP} that relies on the associated video stream of an audio-visual recording capturing the lip movements corresponding to the spoken utterance of the target speaker. However, uni-modal \ac{TSE} systems struggle to extract the target speaker accurately when the auxiliary information provided is unreliable.

\Ac{MTSE} systems take advantage of complementary information from different modalities, thus avoiding the shortcomings of uni-modal systems. In this work, we use the enrolment signal and the video stream as auxiliary information~\cite{Ochiai2019,afouras2019my, Hiro_multimodal_2021}. Existing \ac{MTSE} systems encounter challenges, such as modality dominance, where one of the modalities dominates the other~\cite{michelsanti2021overview}, and modality dropout when one of the modalities is unavailable. A practical \ac{MTSE} system should perform well when only one of the modalities is available or reliable. Additionally, it should be robust to using extracted speech as an enrollment signal (i.e., self-enrolment) when the target speaker is not enrolled and the video stream is unreliable. 

Robustness to modality dropout was addressed using \ac{MTT} in~\cite{Ochiai2019}.
Sato et al.~\cite{Hiro_multimodal_2021} introduced an attention scheme for combining the auxiliary information from different modalities and modality-aware training to improve the system's robustness against unreliable input modalities. Afouras~et~al.~\cite{afouras2019my} addressed the robustness to partial occlusions in the video stream using data augmentation. However, none of these works comprehensively evaluated the performance for all different inference conditions encountered by a practical \ac{MTSE} system. Moreover, these works only investigated non-causal systems. %, which cannot be used in a real-time communication scenario. %In contrast to previous studies that focused solely on non-causal systems, we also investigate the performance of causal systems, which are required for real-time communication scenarios. %In contrast to previous studies that focused solely on non-causal systems, we also investigate the performance of causal systems, which are required for real-time communication scenarios.
%Unlike these studies  which considered only non-causal systems, we investigate the performance of a causal system which is essential for real-time communication scenarios.  

The present study evaluates several existing strategies to train an \ac{MTSE} system that is robust against modality dropout. Our study includes the usage of a simple training strategy, referred to as \ac{MDT}, which applies dropout to the auxiliary information streams in each training step. % to develop a robust \ac{MTSE} system. 
%and compare its performance with the standard and \ac{MTT}~\cite{Ochiai2019} training strategies. 
Although, \ac{MDT} has been applied successfully in other domains~\cite{Neverova_ModDrop_2016,Abdelaziz_2020_ICMI_ModDrop,Saghir_2022_CVPR_ModDrop,omalley22_interspeech}
%such as gesture recognition~\cite{Neverova_ModDrop_2016}, driving animated faces~\cite{Abdelaziz_2020_ICMI_ModDrop}, action recognition~\cite{Saghir_2022_CVPR_ModDrop} and universal front-ends for \ac{ASR}~\cite{omalley22_interspeech}, 
its utilization within the context of \acl{MTSE} remains unexplored. In contrast to the aforementioned works that solely investigated non-causal systems, we also investigate the performance of causal systems, which are required for real-time communication scenarios. Our experiments demonstrate that the system trained with \ac{MDT} has reduced sensitivity to architectural choices, notably the normalization layers in the \ac{DNN}, compared to existing training strategies in non-causal and causal configurations. We also show that a system trained with \ac{MDT} avoids modality dominance compared to systems trained using other strategies. In addition, we investigate the system performance with self-enrolment ~\cite{Ge_Spex++_2021,afouras2019my} and show that the system trained with \ac{MDT} performs better than systems trained with other strategies when utilizing extracted speech as enrolment signal in non-causal and causal configurations.

\begin{figure*}[t]
    \centering
    \includegraphics[width=0.9\linewidth]{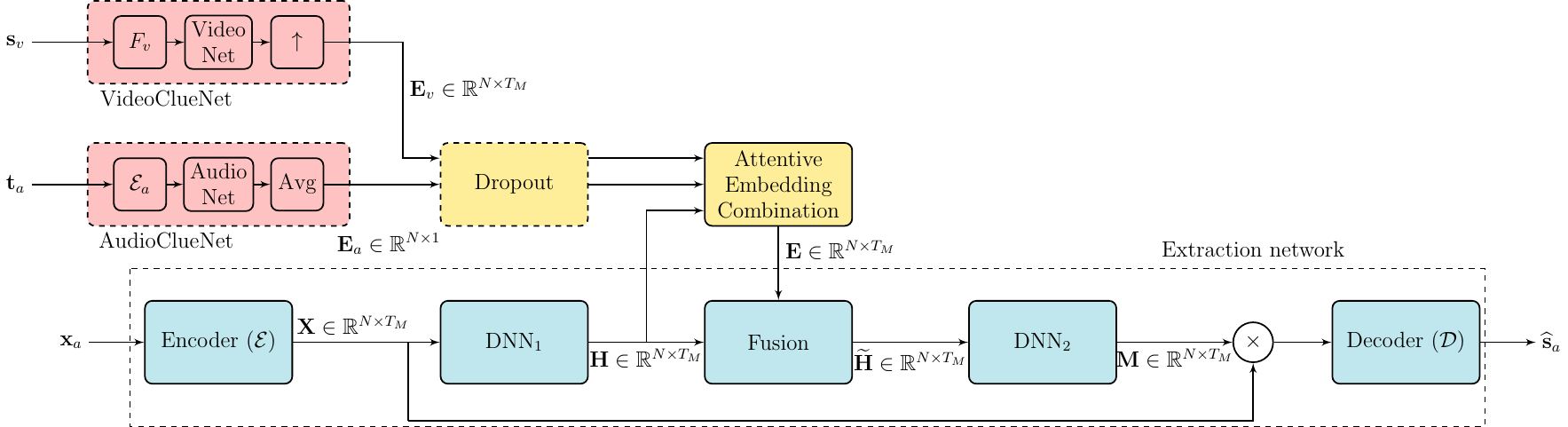} % columnwidth
    \caption{The \acf{MTSE} system under test. The \acf{MDT} strategy is only applied during training. %The dropout block is removed for training with \acf{MTT} and standard (without dropout) training strategies.
    }
    \label{fig:block_diagram}
\end{figure*}

\section{Problem formulation}\label{sec:prob_form}
\vspace{-2mm}
Consider the scenario of an audio-visual recording of a single speaker of interest (i.e., the target speaker) and an interfering speaker. In this work, we assume the interference to be from a different speaker and is present only in the
input mixture audio, and the video stream is assumed to contain only the target speaker. In addition, an enrolment utterance of the target speaker is assumed to be available. The goal of an \ac{MTSE} system is to extract the target speaker's speech from the input speech mixture based on the associated video stream and audio enrolment.

Formally, let ${\bf s_\textrm{a}} \in \mathbb{R}^{T \times 1}$ and ${\bf s}_\textrm{v}\in \mathbb{R}^{H \times W \times T_\textrm{v}}$ represent the audio signal and video stream corresponding to the target speaker. Here, $T$ denotes the number of time samples, $H$, $W$ and $T_\textrm{v}$ indicates the height, width and the number of video frames, respectively. Let ${\bf i_\textrm{a}} \in \mathbb{R}^{T \times 1}$ denote the interfering speaker, ${\bf x_\textrm{a}}$ is the input audio mixture, and ${\bf t_\textrm{a}} \in \mathbb{R}^{T_\textrm{a} \times 1}$ indicate the target speaker audio enrolment with duration $T_\textrm{a}$. The goal of the \ac{MTSE} system is to obtain an estimate of ${\bf s_\textrm{a}}$ given the input mixture ${\bf x_\textrm{a}}= {\bf s_\textrm{a}} + {\bf i_\textrm{a}}$, the enrolment utterance ${\bf t_\textrm{a}}$, and the video stream ${\bf s}_\textrm{v}$, i.e.,
\begin{equation}\label{eqn:mtse_objective}
    \hat{\bf s}_\textrm{a} = \textrm{MTSE}\left({\bf x_\textrm{a}} | {\bf s}_\textrm{v}, {\bf t_\textrm{a}} \right).
\end{equation}
As in previous works~\cite{Ochiai2019, afouras2019my, Hiro_multimodal_2021}, a data-driven approach is employed to design the MTSE system, which is realized using a \ac{DNN}.

\section{Multi-modal Target Speaker Extraction}\label{sec:proposed}
\vspace{-2mm}
Figure~\ref{fig:block_diagram} shows a block diagram of the \ac{MTSE} system used in this study. The architecture comprises an AudioClueNet module, a VideoClueNet module, an embedding combination module, and an extraction network. 

The extraction network uses ''masking in the learned time-feature domain'' \cite{Luo2018} strategy in which the input speech mixture ${\bf x_{a}}$ is first fed to a time-feature encoder $\mathcal{E}$ to compute sub-sampled features ${\bf X}$, which are then fed to the block $\textrm{DNN}_1$ to compute intermediate representations ${\bf H} \in \mathbb{R}^{N \times T_{M}}$, where $N$ denotes the number of channels and $T_{M}$ represents the number of time frames. The representation ${\bf H}$ is then fused with the embedding ${\bf E}$ derived by combining the embeddings extracted from the audio and the video auxiliary information.

The enrolment audio ${\bf t}_\textrm{a}$ is processed by an AudioClueNet, composed of a time-feature encoder $\mathcal{E}_{a}$, an AudioNet module, and temporal averaging, to compute a \acl{TI} embedding ${\bf E}_\textrm{a} \in \mathbb{R}^{N \times 1}$. The embedding ${\bf E}_\textrm{a}$ captures speaker-specific characteristics and hence informs the extraction network about the target speaker. 

Since the audio of the target speaker in the input speech mixture and the associated video stream are temporally correlated as they are recorded in parallel, the VideoClueNet extracts \acl{TV} embeddings, one per each input frame of the video stream ${\bf s}_\textrm{v}$. This is achieved by passing ${\bf s}_\textrm{v}$ through a visual front-end $F_\textrm{v}$, a VideoNet module, and an upsampling block to obtain ${\bf E}_\textrm{v} \in \mathbb{R}^{N \times T_{M}}$ whose dimension matches with that of the embedding ${\bf H}$, i.e.,
\begin{equation}
    {\bf E}_\textrm{v} = \textrm{Upsample}(\textrm{VideoNet}(F_\textrm{v}({\bf s}_\textrm{v}))) \in \mathbb{R}^{N \times T_{M}}.
\end{equation}
The upsampling block is realized by linear interpolation. The embeddings from the AudioClueNet and VideoClueNet are then combined using the attentive combination function $f$ described in Sec.~\ref{sec:aec} to derive the joint embeddings ${\bf E}$. 

The combined embeddings ${\bf E} = \{ {\bf E}_1, {\bf E}_2, \ldots, {\bf E}_{T_M}\}$ are then fused with the mixture representations ${\bf H}$ via element-wise multiplication similar to~\cite{Ochiai2019} in the fusion block, i.e.,
\begin{equation}
    \widetilde{\bf H} = {\bf H}  \odot {\bf E}.
\end{equation}
We chose element-wise multiplication as a fusion strategy since it performed similarly to sum and concatenation-based methods~\cite{Ochiai2019, Hiro_multimodal_2021}.
The modified representations $\widetilde{\bf H}$ are then fed to the second \ac{DNN} block ($\textrm{DNN}_2$), which generates the mask ${\bf M}$. The masked mixture features are then fed to the learned decoder to compute the extracted signal $\hat{\bf s}_\textrm{a}$,
\begin{equation}
    \hat{\bf s}_\textrm{a} = \mathcal{D}( {\bf X} \odot {\bf M}).
\end{equation}

%The AudioClueNet, which comprises a time-feature encoder $\mathcal{E}_{a}$, AudioNet, and averaging blocks, is used to compute a \ac{TI} embedding ${\bf E}_\textrm{a} \in \mathbb{R}^{N \times 1}$ from the enrolment utterance ${\bf t}_\textrm{a}$.

%The embedding ${\bf E}_\textrm{a}$ is obtained by first passing the enrolment sequence through time-feature encoder $\mathcal{E}$ followed by 
%The enrolment utterance is pre-recorded %and often recorded at a different time. 
%and represents the target speaker's characteristics. The system's performance can be sensitive to the choice of enrolment utterance, its duration, and real applications often encounter users who are not enrolled.  
%However, in practical scenarios, the video stream is often corrupted by occlusions, and in communication applications, the video frames may be dropped due to unreliable communication bandwidth, or the target speaker may turn the camera off. Furthermore, on resource-constrained devices, the computational requirements for the VideoClueNet may be prohibitive for real-time operation since it typically has more parameters than the audio extraction network itself.

\subsection{Attentive embedding combination}\label{sec:aec}
% \vspace{-2mm}
We adopt the attentive embedding combination scheme proposed in~ \cite{Ochiai2019, Hiro_multimodal_2021} to combine the audio and video embeddings. In this scheme, the combined embedding at time frame $t$, ${\bf E}_t$, is obtained as a convex combination of the individual embeddings, 
\begin{equation}
    \label{eq:weavg}
      {\bf E}_t = f({\bf E}_\textrm{a}, {\bf E}_{\textrm{v},t}) \triangleq w_{\textrm{a},t}\,{\bf E}_\textrm{a} + w_{\textrm{v},t}\,{\bf E}_{\textrm{v},t}.
\end{equation}
The weights $\{w_{\textrm{a},t}, w_{\textrm{v},t}\}$ are computed by cross-attention with the mixture signal hidden representation ${\bf H}_t$ at time $t$ to emphasize the modality that is important to extract the target speaker from the mixture at the current time frame, 
\begin{eqnarray}
      e_{\textrm{a},t} &=& {\bf w}^\textrm{T}~\mbox{tanh}({\bf W} \,{\bf H}_t + {\bf V} \, {\bf E}_{\textrm{a}} + {\bf b}), \\ 
      e_{\textrm{v},t} &=& {\bf w}^\textrm{T}~\mbox{tanh}({\bf W} \, {\bf H}_t + {\bf V} \, {\bf E}_{\textrm{v},t} + {\bf b}), \\ 
      w_{q,t} &=& \frac{\exp{(\gamma\, e_{q,t}})}{ \exp{(\gamma\, e_{\textrm{a},t}}) + \exp{(\gamma\, e_{\textrm{v},t}})},~~q \in \{\textrm{a},\textrm{v}\},
\label{eq:e_theta}
\end{eqnarray}
where $\gamma$ is a sharpening factor, and $\{{\bf w}\in \mathbb{R}^{N \times 1}, {\bf W}\in \mathbb{R}^{N \times N}, {\bf V}\in \mathbb{R}^{N \times N}, {\bf b}\in \mathbb{R}^{N \times 1} \}$ are trainable parameters. 
%The normalization in \eqref{eq:e_theta} ensures that the combined embedding ${\bf E}_t$ is a convex combination of the individual modality embeddings ${\bf E}_\textrm{a}$ and ${\bf E}_{\textrm{v},t}$.

\section{Training Strategies} \label{sec:training_strategy}
% \vspace{-2mm}
The \ac{MTSE} model described in Sec.~\ref{sec:proposed} was trained using three different training strategies to study their effect on the model's robustness. 

\subsection{Standard training (ST)}\label{ss:standard}
% \vspace{-2mm}
In the standard training strategy, the model is trained with the \ac{SI-SDR}~\cite{leroux2019} as a loss function by feeding the model with the mixture signal, along with the audio and video auxiliary information, during the forward pass. 
%SI-SDR is used as the loss function
%The without the dropout strategy. 

\subsection{Multi-task training (MTT)}\label{ss:MTT}
% \vspace{-2mm}
%The multi-task loss~\cite{Ochiai2019} consists of three components: a) loss component $L_\textrm{av}$ with both auxiliary information available. b) loss component $L_\textrm{a}$ with only audio auxiliary information available. c) loss component $L_\textrm{v}$ with only video auxiliary information available. 
In the \ac{MTT} strategy, as proposed in \cite{Ochiai2019}, each example during training is fed three times to the model, first with both auxiliary information and then with audio-only and video-only auxiliary information. The three extracted signals are then used to compute three loss terms $L_\textrm{av}$, $L_\textrm{a}$ and $L_\textrm{v}$ respectively using \ac{SI-SDR} as the loss function. A weighted sum of these loss terms is used as the final loss term $L_\textrm{MTT}$ for the back-propagation training. In this study,~\ac{MTT} loss ($L_\textrm{MTT}$) is computed by equally weighing the loss terms as:
% , using the weights suggested in \cite{Ochiai2019}
\begin{equation}\label{eqn:MTT}
L_\textrm{MTT} = \frac{1}{3}\, L_\textrm{av} + \frac{1}{3}\, L_\textrm{a} + \frac{1}{3}\, L_\textrm{v}. \vspace{-0.25em}
\end{equation}

%The weights used for the loss terms are the same as those proposed in~\cite{Ochiai2019}. 

\subsection{Modality dropout training (MDT)}\label{ss:dropout}
% \vspace{-2mm}
%In this strategy, 
Inspired by~\cite{Neverova_ModDrop_2016, Abdelaziz_2020_ICMI_ModDrop, Saghir_2022_CVPR_ModDrop,omalley22_interspeech}, in the \ac{MDT} strategy, the model is trained with the \ac{SI-SDR} loss by passing the embedding ${\bf E}$ from the audio and video auxiliary information along with the mixture signal. The embedding ${\bf E}$ is computed as,
\vspace{-1mm}
\begin{equation}\label{eqn:drop}
    {\bf E} = \begin{cases}
                f( {\bf E}_\textrm{v}, {\bf E}_\textrm{a} )&w. p.~1/3\\
                f( {\bf E}_\textrm{v}, {\bf 0} )&w. p.~1/3\\
                f( {\bf 0}, {\bf E}_\textrm{a} )&w. p.~1/3\\
                \end{cases}.
\end{equation}
Consequently, for all the training samples within the batch, with a probability of $1/3$, all auxiliary information or only one of the auxiliary information are used for training. It is ensured that both auxiliary information are not dropped simultaneously. When an auxiliary information is dropped, its corresponding embedding is replaced with an all-zero vector/matrix. This scheme of using an all-zero embedding vector for a dropped modality has the advantage that the corresponding embedding extraction network can be bypassed during inference, which saves computational resources.

\vspace{-0.5em}
\section{Experimental Setup}\label{sec:exp_setup}
% \vspace{-2mm}
\subsection{Model description}\label{model_decscription}
% \vspace{-2mm}
The basic building block of \ac{MTSE} system under test is the \ac{DPRNN} proposed in~\cite{Luo_2020_DPRNN}. The time-feature encoders $\mathcal{E}$ and $\mathcal{E}_{a}$ are realized by a 1D convolutional layer with 2~\unit{ms} kernel size, 1~\unit{ms} stride and 256 channels. The visual front-end $F_\textrm{v}$~\cite{Stafylakis2017} is pre-trained and extracts a $512$-dimensional visual feature from the lip region extracted from each frame in the video stream. This visual front-end consists of a 3D convolutional layer followed by four blocks of ResNet-18~\cite{he2015deep}. The blocks AudioNet, VideoNet, $\textrm{DNN}_1$, and $\textrm{DNN}_2$ comprise of \ac{DPRNN} blocks with the following hyperparameters: chunk size $K=100$, two DPRNN layers in each block, and the hidden dimension was set to 128. A DPRNN layer consists of a cascade of inter-chunk and intra-chunk RNNs. 
The inter- and intra-chunk RNNs are realized in the non-causal configuration using bi-directional \ac{LSTM}. However, for the causal configuration, inter-RNN uses a uni-directional \ac{LSTM}, and the intra-RNN uses a bi-directional \ac{LSTM}. A 1$\times$1 convolutional layer is used in the VideoNet to match the dimension of the outputs of $F_\textrm{v}$ and \ac{DPRNN} blocks. The extraction network's decoder $\mathcal{D}$ consists of a 1D transpose convolution layer with the same kernel size and hop size as the encoder $\mathcal{E}$.

%The output of the learnable encoder is then fed to \ac{DPRNN} blocks. The VideoClueNet uses a pre-trained visual front-end~\cite{Stafylakis2017} to extract $512$-dimensional visual features from the lip region before feeding them to a 1$\times$1 convolutional layer and \ac{DPRNN} block. This visual front-end consisted of a 3D convolutional layer followed by four blocks of ResNet-18~\cite{he2015deep}. The $\textrm{DNN}_2$ block in the extraction network also comprises of \ac{DPRNN} blocks. All the \ac{DPRNN} blocks share the same hyperparameters. The chunk size $K$ is set to 100, whereas the number of blocks used is 2. The \ac{DPRNN} blocks use bi-directional \ac{LSTM} as inter- and intra-chunk RNNs with a hidden dimension of 128. However, for the causal configuration, inter-RNN uses a uni-directional \ac{LSTM}. The decoder of the main network consisted of a 1D transpose convolution layer with the same kernel size and hop size as the encoder.

%The model under test consists of $15.12$~M parameters, out of which the VideoClueNet has $12.23$~M parameters, whereas the extraction network and the AudioClueNet has only $1.84$~M and $0.92$~M parameters, respectively.
%processing the input mixture ${\bf x}_\textrm{a}$

\subsection{Datasets and training}\label{datasets_training}
% \vspace{-2mm}
For our experiments, we generated two-speaker mixtures from the LRS3 dataset, a large-scale audio-visual dataset obtained from TED and TEDx talks \cite{afouras2018lrs3}. The dataset was created similar to~\cite{elmin2022newinsights} where the target and interferer speech were randomly selected from different speaker identities and mixed with a \ac{SIR} randomly sampled from $-5$~dB to $5$~dB. Speaker identities that have less than three audio samples were excluded from the dataset creation. The training, validation, and test splits of the dataset consisted of 20k, 4k, and 2k samples, respectively. All audio samples had a $16$~\unit{kHz} sampling rate and a $3$~\unit{s} duration. The video clips were preprocessed before feeding to the VideoClueNet such that only the lip regions were extracted, converted to greyscale, and scaled to a dimension of $100\times 50\times1$ corresponding to width $W$, height $H$, and channel, respectively. The enrolment audio had a duration of $3$~\unit{s}.

For training the models, we used Adam optimizer~\cite{Kingma2015} with initial learning rate $5\times10^{-4}$ and a weight decay~$10^{-5}$. For the models trained with standard and with \ac{MDT} strategy, the batch size was~$20$. The batch size was set to $7$ for the \ac{MTT} as it involves three forward passes through the model. Gradient clipping was employed when the $\ell_{2}$-norm exceeded the value~$5$. The models were trained for a maximum of $300$~epochs with a scheduler that reduced the learning rate if the validation loss did not reduce for $5$~consecutive epochs. Early stopping with a patience of $40$ epochs was used. Dynamic mixing~\cite{zeghidour2021wavesplit} was employed to ensure that the training samples differed in each epoch. The sharpening factor $\gamma$ in (\ref{eq:e_theta}) was set to 2 as in~\cite{Hiro_multimodal_2021}.

%For the model trained with \ac{MTT}, batch size was set to $7$ due to memory constraints. Gradient clipping was employed when the $\ell_{2}$-norm exceeded the value~$5$. The models were trained for a maximum of $300$~epochs with a scheduler that reduced the learning rate if the validation loss did not reduce for $5$~consecutive epochs. Early stopping was employed with the patience of $40$ epochs. Dynamic mixing~\cite{zeghidour2021wavesplit} was employed to ensure that the training samples differed in each epoch. The sharpening factor $\gamma$ is set to 2 as in~\cite{Hiro_multimodal_2021}.

\subsection{Conditions during inference}\label{ss:inference_modes_of_operation}
\vspace{-2mm}
During inference, the \ac{MTSE} systems were evaluated under four inference conditions, which are of practical importance as described in Sec.~\ref{sec:intro}:
\begin{itemize}
    \item~\textbf{\ac{MTSE}}: In this condition, both the audio and video modalities are assumed to be available i.e., ${\bf E}_\textrm{a} = \textrm{AudioClueNet}({\bf t_\textrm{a}})$, ${\bf E}_\textrm{v} = \textrm{VideoClueNet}({\bf s}_\textrm{v})$.

    \item~\textbf{\ac{AoTSE}}: In this condition, the video modality is assumed to be unavailable, but the audio modality is assumed to be available. For system trained with \ac{MTT}, $\bf s_\textrm{v} = {\bf 0}_{H \times \times W \times T_\textrm{v}}$, and for system trained with \ac{MDT}, ${\bf E}_\textrm{v} = {\bf 0}_{N \times T_M}$.

    \item~\textbf{\ac{VoTSE}}: In this condition, we consider the scenario where the target speaker is not enrolled, but the video modality is assumed to be available. In this case, the audio embedding is a vector of zeros, ${\bf E}_\textrm{a} = {\bf 0}_{N \times 1}$ for systems trained with \ac{MDT}. For systems trained with \ac {MTT}, ${\bf s}_\textrm{a} = {\bf 0}_{T \times 1}$.

    \item~\textbf{MTSE-FD}: Here, we consider the scenario where the video frames have dropped frames due to occlusion or low communication bandwidth. We consider burst frame drops at a rate of $1/3$ for our evaluation.
    %For the dropped frames, the corresponding video embedding ${\bf E}_\textrm{v,t}$ is set to a vector of zeros.  
\end{itemize}

\section{Results} \label{sec:results}
\vspace{-2mm}
 %was used as the evaluation metric. The systems compared are:
We evaluated the uni-modal and multi-modal \ac{TSE} systems in non-causal and causal configurations. Our initial experimentation showed the models to be sensitive to the normalization layers used in the \ac{DNN} architecture. Hence, we study the choice of the normalization layer in addition to training strategies. In particular, \ac{gLN}~\cite{Luo2019} and \ac{LN}~\cite{ba2016layer} are investigated for the non-causal configuration,  and for the causal configuration, \ac{cLN}~\cite{Luo2019} and \ac{LN} are investigated. During our study, we observed that the non-causal~\ac{MTSE} system trained with the~\ac{MTT} strategy with weights as proposed in~\cite{Ochiai2019} and~\ac{gLN} completely failed for the \ac{AoTSE} condition. Therefore, equal weights as proposed in Sec.~\ref{ss:MTT} were used. The \ac{MTSE} models were trained using the strategies explained in Sec.~\ref{sec:training_strategy}. The performance of these systems was evaluated using the commonly used \ac{SI-SDR} metric~\cite{leroux2019}. 

  \begin{table}[t]
    \centering
    \setlength{\tabcolsep}{7pt}
    \renewcommand{\arraystretch}{1.5}%%%
    \caption{SI-SDR improvement [dB] (mean and \acl{SD}) for the evaluated systems. %We dropped the video frames for burst frame drops corresponding to a frame drop rate of $33.33\%$. 
    %$\textrm{U-ST}$ and $\textrm{M-ST}$ are uni-modal and multi-modal systems trained using the standard training (ST) strategy respectively. \ac{MTT} and \ac{MDT} training strategies are used for training only the multi-modal systems. 
    For reference, the performance of the uni-modal non-causal (causal) AoTSE is \mbox{12.9 $\pm$ 5.5} (\mbox{12.2 $\pm$ 5.3}), and for VoTSE is \mbox{13.9 $\pm$ 3.3} (\mbox{13.5 $\pm$ 3.5}).
    }
    %\vspace{-0.5em}
    \scalebox{0.75}
    {
 %\begin{tabular}{@{}llc{S}{S}{S}{S}@{}}
 \begin{tabular}{@{}llcSSSS@{}}
 \toprule
  & \textbf{Tr. Strategy} & \textbf{Norm} & \textbf{MTSE} & \textbf{AoTSE} & \textbf{VoTSE} & \textbf{MTSE-FD}  \\ %Attention
 \midrule
 %\multirow{8}{*}{\rotatebox[origin=c]{90}{\textbf{non-causal}}} & U-ST & gLN & {-} & {12.9 $\pm$ 5.5} & {-} & {-}   \\ %- &
 %& U^{st} & gLN & {-} & {-} & {13.9 $\pm$ 3.3} & {-} \\ %& -
 \multirow{6}{*}{\rotatebox[origin=c]{90}{\textbf{non-causal}}} & ST & gLN & {14.3 $\pm$ 3.1} & {11.1 $\pm$ 7.9} & {10.5 $\pm$ 4.2} & {13.5 $\pm$ 4.6}  \\ %& SA
 & ST & LN & {14.5 $\pm$ 3.1} & {3.3 $\pm$ 7.1} & {9.4 $\pm$ 3.5} & {13.4 $\pm$ 5.0}  \\ %& SA
 %MM-NA & No & No & xxx & - & - & - \\ % & NA
 %& MTT~\cite{Ochiai2019}  & gLN & {14.5 $\pm$ 3.6} & {{$-$}7.4 $\pm$ 20.1} & {14.5 $\pm$ 3.6} & {12.5 $\pm$ 6.9} \\
 & MTT  & gLN & {14.5 $\pm$ 3.3} & {10.9 $\pm$ 10.3} & {14.4 $\pm$ 3.2} & {13.9 $\pm$ 4.3} \\
 %~\cite{Ochiai2019}
 %MM-NA & No & No & xxx & - & - & - \\ % & NA
 %& MTT~\cite{Ochiai2019}  & LN & {14.5 $\pm$ 3.5} & {11.6 $\pm$ 9.5} & {13.4 $\pm$ 3.5} & {14.0 $\pm$ 4.6} \\
  & MTT  & LN & {14.6 $\pm$ 3.3} & {12.7 $\pm$ 7.6} & {14.5 $\pm$ 3.4} & {14.2 $\pm$ 4.2} \\
 & MDT & gLN & {13.9 $\pm$ 3.5} & {12.8 $\pm$ 6.3} & {13.9 $\pm$ 3.3} & {13.2 $\pm$ 4.8} \\ %& SA
 & MDT & LN & {13.9 $\pm$ 3.4} & {12.8 $\pm$ 6.5} & {13.9 $\pm$ 3.2} & {13.4 $\pm$ 4.6} \\ %& SA
 \midrule
 %\multirow{8}{*}{\rotatebox[origin=c]{90}{\textbf{causal}}} & U^{st} & LN & {-} & {12.2 $\pm$ 5.3} & {-} & {-}  \\ %& - 
 %& U^{st} & LN & {-} & {-} & {13.5 $\pm$ 3.5} & {-}   \\ 
 %MM-Baseline & No & Yes & bN & 13.73 & 11.84 & -1.97 & 11.72 & 1.65 \\ %& SA
 \multirow{6}{*}{\rotatebox[origin=c]{90}{\textbf{causal}}} & ST & cLN & {13.2 $\pm$ 3.8} & {7.5 $\pm$ 7.0} & {9.4 $\pm$ 4.8} & {11.3 $\pm$ 6.0} \\ %& SA
 & ST & LN & {13.9 $\pm$ 3.4} & {7.4 $\pm$ 7.6} & {8.6 $\pm$ 4.3} & {12.6 $\pm$ 5.4} \\ %& SA
 %MM-NA & No & Yes & xxx & - & - & - \\ % & NA 
 %MTT & No & Yes & bN & 13.72 & 7.55 & 11.70 & 12.13 & 10.48 \\
 %& MTT~\cite{Ochiai2019} & cLN & {13.7 $\pm$ 3.9} & {8.6 $\pm$ 9.7} & {12.3 $\pm$ 5.0} & {11.7 $\pm$ 6.2}  \\
 %& MTT~\cite{Ochiai2019}  & LN & {14.4 $\pm$ 3.3} & {11.1 $\pm$ 9.7} & {14.0 $\pm$ 3.7} & {13.4 $\pm$ 5.0} \\
  & MTT & cLN & {13.6 $\pm$ 3.7} & {8.4 $\pm$ 9.5} & {12.5 $\pm$ 4.5} & {12.3 $\pm$ 5.5}  \\
 & MTT  & LN & {14.3 $\pm$ 3.3} & {11.6 $\pm$ 9.3} & {14.1 $\pm$ 3.6} & {13.7 $\pm$ 4.5} \\
%Proposed & Yes & Yes & bN & 11.20 & 10.08 & 7.92 & 10.92 & 9.03  \\ %& SA 
 %MM-NA & No & Yes & xxx & - & - & - \\ % & NA 
 & MDT & cLN & {13.2 $\pm$ 3.7} & {12.2 $\pm$ 6.0} & {12.3 $\pm$ 4.4} & {11.9 $\pm$ 5.5}  \\ %& SA 
 %MM-NA & No & Yes & xxx & - & - & - \\ % & NA 
 & MDT & LN & {13.4 $\pm$ 3.3} & {12.5 $\pm$ 5.9} & {13.3 $\pm$ 3.6} & {12.7 $\pm$ 4.7}  \\ %& SA 
 \bottomrule
    \end{tabular}
    }
    \label{tab:results}
%\vspace{-4mm}
\end{table} 

%\vspace*{-\baselineskip}

%\vspace{-4mm}
\subsection{Performance for different inference conditions} %layer}
\label{sec:inf_modes}
%\vspace{-2mm}
Table~\ref{tab:results} shows the results in terms of SI-SDR improvement. In the non-causal configuration, for the \ac{MTSE} condition, the standard and \ac{MTT} strategies perform better than the \ac{MDT} strategy in terms of the mean. However, the standard training strategy has a smaller \ac{SD} compared to both \ac{MTT} and \ac{MDT} strategies. Both standard and \ac{MTT} strategies are found to be sensitive to the choice of the normalization layer in \ac{AoTSE} and \ac{VoTSE} conditions. The performance of the model using \ac{LN} and trained with the standard training strategy is found to degrade significantly in the \ac{AoTSE} and \ac{VoTSE} conditions. %We observe that the model using \ac{gLN} and trained with \ac{MTT} strategy completely fails for the \ac{AoTSE} condition. This clearly shows that the system suffers from modality dominance, where the video modality dominates the audio modality. In contrast, the proposed \ac{MDT} strategy exhibits robustness (both in terms of mean and \ac{SD}) to the choice of the normalization layer across all the inference conditions. 
Furthermore, the \ac{MDT} strategy is less prone to the modality dominance, and the performance for the \ac{AoTSE} and \ac{VoTSE} conditions is also close to the corresponding uni-modal \ac{TSE} systems. For both \ac{MTT} and \ac{MDT} strategies, the \ac{MTSE} system using \ac{LN} exhibits robust performance to missing video frames (i.e., MTSE-FD condition), where a slight drop in performance can be seen compared to the \ac{MTSE} condition.

% in the MTSE-FD condition (i.e., dropped video frames), a slight drop in performance can be seen for all training strategies.  
% In the MTSE-FD condition, the performance is better than \ac{AoTSE} for all systems and approaches that of \ac{MTSE} in terms of mean but exhibits a higher \ac{SD}. 

In the causal configuration, we observed that using \ac{LN} instead of  \ac{cLN} generally improves the performance for all inference conditions, except for the standard training in \ac{AoTSE} and \ac{VoTSE} conditions. 
For the system trained with standard training strategy, a significant drop in performance can be seen in the \ac{AoTSE} and \ac{VoTSE} conditions. Similar to non-causal configuration, the system trained with \ac{MTT} strategy suffers from dominance of the video modality. 
It can also be observed that the system trained with the \ac{MDT} strategy is less prone to modality dominance. In addition, the performance is also comparable to the corresponding uni-modal systems, albeit with slightly higher \ac{SD}. % However, for the system trained with standard training strategy, there is a drop in terms of mean albeit smaller \ac{SD}.  

\subsection{Self-enrolment}\label{sec:sim_scenario}

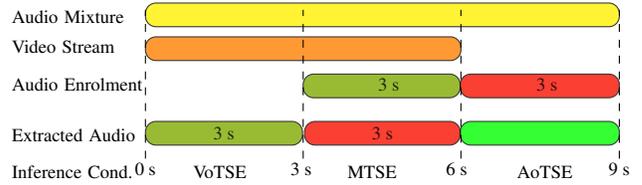
\begin{figure}[t]
    \centering
    \captionsetup{aboveskip=10pt}
    \vspace{-1.5em}
    \begin{tikzpicture}[font=\scriptsize, scale=0.9]
% Rectangle 4
\draw[fill=green, opacity=0.8, rounded corners] (4.65,0.1) rectangle (7,0.45) node[midway] {};
\draw[fill={rgb:red,128;green,4;yellow,5}, opacity=0.8, rounded corners] (2.35,0.1) rectangle (4.65,0.45) node[midway] {3 \unit{s}};
\draw[fill={rgb:red,2;green,3;yellow,1}, opacity=0.8, rounded corners] (0,0.1) rectangle (2.32,0.45) node[midway] {3 \unit{s}};
% Rectangle 3
\draw[fill={rgb:red,128;green,4;yellow,5}, opacity=0.8, rounded corners] (4.65,0.8) rectangle (7,1.15) node[pos=0.545] {3 \unit{s}};
%\draw[fill={rgb:red,2;green,3;yellow,1}, opacity=0.8, rounded corners] (4.65,0.8) rectangle (5.8,1.15) node[pos=0.545] {3 \unit{s}};
\draw[fill={rgb:red,2;green,3;yellow,1}, opacity=0.8, rounded corners] (2.33,0.8) rectangle (4.65,1.15) node[pos=0.545] {3 \unit{s}};
% Rectangle 2
\draw[fill=orange, opacity=0.8, rounded corners] (0,1.35) rectangle (4.66,1.7) node[pos=0.51] {};
% Rectangle 1
\draw[fill=yellow, opacity=0.8, rounded corners] (0,1.85) rectangle (7,2.2) node[midway] {};

% Label at specified position
\node[text width=2.2cm] at (-0.75, 0.25) {Extracted Audio} ;
\node[text width=2.2cm] at (-0.75, 1,0) {Audio Enrolment};
\node[text width=2.2cm] at (-0.75, 1.55) {Video Stream};
\node[text width=2.2cm] at (-0.75, 2.0) {Audio Mixture};
\node[text width=2.2cm] at (-0.75, -0.3) {Inference Cond.};
% Vertical dashed lines dividing rectangles equally
\draw[dashed] (0,-0.1) -- (0,2.1);
\draw[dashed] (2.33,-0.1) -- (2.33,2.1);
\draw[dashed] (4.66,-0.1) -- (4.66,2.1);
\draw[dashed] (7,-0.1) -- (7,2.1);

% Nodes with specified positions
\node at (0,-0.25) {0 \unit{s}};
\node at (2.3,-0.25) {3 \unit{s}};
\node at (4.6,-0.25) {6 \unit{s}};
\node at (7,-0.25) {9 \unit{s}};

% Label at specified position
\node at (1.1, -0.3) {VoTSE};
\node at (3.35, -0.3) {MTSE};
\node at (5.9, -0.3) {AoTSE};

% Curved arrow
%\draw[->, bend left=40] (1.1,0.65) to[out=2, in=-150] (3.35,0.9);
%\draw[->, bend left=40] (1.1,0.35) to[out=5, in=-160] (5.9,1.0);
%\draw[->, bend left=40] (3,0.35) to[out=10, in=-160] (5.9,1.0);

%\draw[<->, dashed, red] (0,0.6) -- (2.3,0.6);
%\node at (1.1, 0.75) {E1};
%\node at (3.4, 0.85) {E1};
%\draw[<->, dashed, red] (2.3,0.9) -- (4.6,0.9);

%\draw[<->, blue] (2.3,0.8) -- (7,0.8);
%\draw[<->, blue] (0,0.5) -- (4.6,0.5);
%\node at (2.5, 0.35) {E2};
%\node at (5.6, 0.85) {E2};

%\draw[->, bend left=40] (1.0,0.6) to[out=2, in=-150] (3.5,0.75);
%\draw[->, bend left=40] (1.1,0.35) to[out=5, in=-160] (5.9,1.0);
\end{tikzpicture}
    % \vspace{-0.5em}
    \caption{Self-enrolment scenario. Extracted audio from the first 3~\unit{s} is used as the enrolment for the next 3~\unit{s}, and the inference in the last 3~\unit{s} is done using the extracted audio up to 6~\unit{s} as the enrolment and the video is assumed to be unavailable.}
    \label{fig:eval_scenario}
    \vspace{1em}
\end{figure}

\begin{figure}[t]
    \centering
    \includegraphics[width=0.9\linewidth]{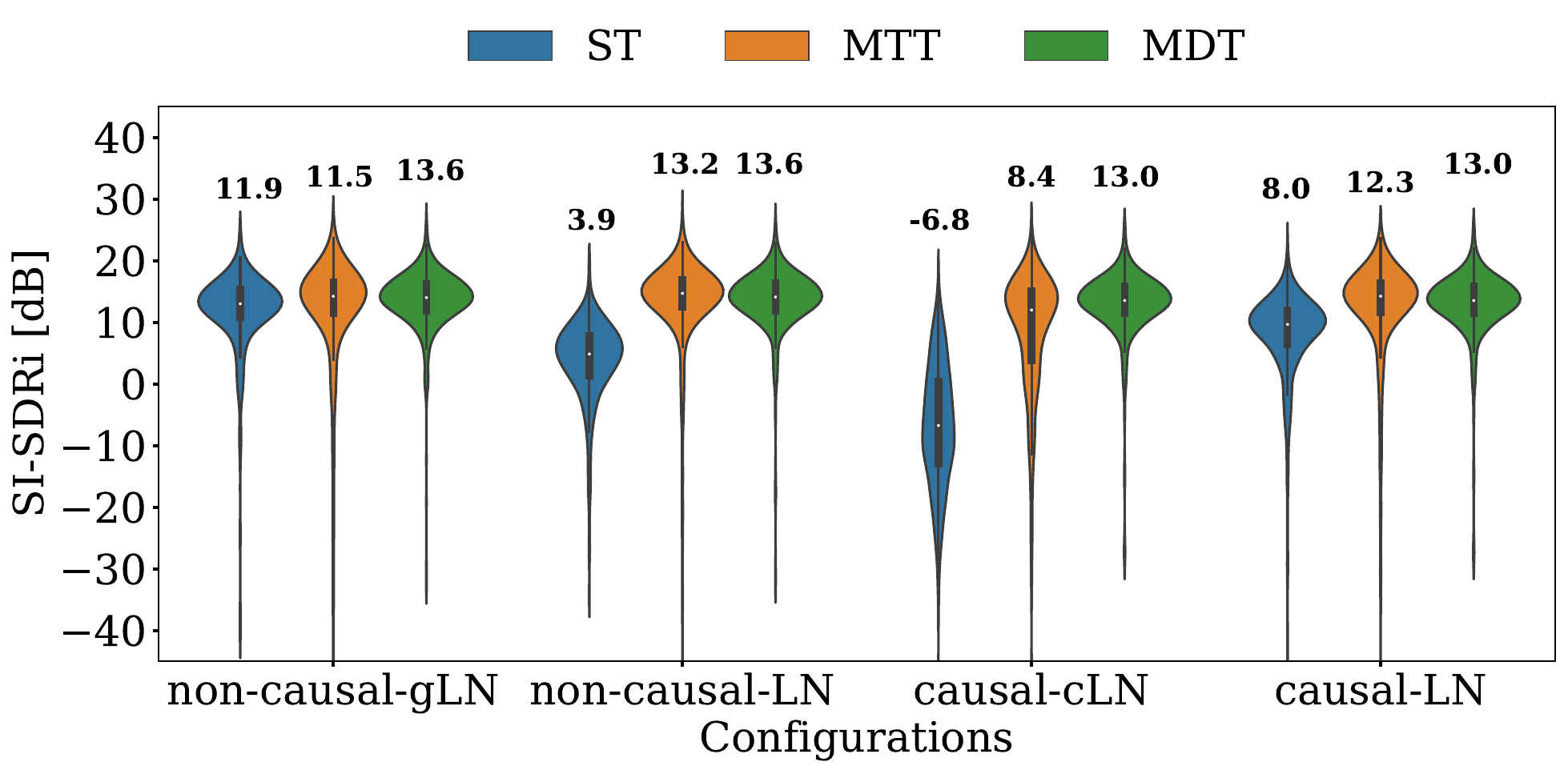}
    \caption{Violin plots showing the distribution of SI-SDR improvements (SI-SDRi) for the 3rd segment of the test examples for different configurations and training strategies. Mean values are shown on top of the distribution.}
    \label{fig:distributions}
    %\vspace{-1.0em}
\end{figure}

In this experiment, we composed 630 audio recordings of 9~\unit{s} from the test splits. Speakers that have less than four audio samples were removed. Figure~\ref{fig:eval_scenario} illustrates the simulated scenario that assumes that the target speaker is not enrolled. For the first 3~\unit{s}, the model operates in \ac{VoTSE} condition since the video stream is assumed to be available and reliable. For the next 3~\unit{s}, the extracted speech from the first 3~\unit{s} is used as the enrolment signal along with the video stream, i.e., the model operated in \ac{MTSE} condition. For the last 3~\unit{s}, the video stream is assumed to be unavailable. However, the extracted speech from the first 6~\unit{s} is used as an enrolment signal, i.e., the model operated in \ac{AoTSE} condition. Figure~\ref{fig:distributions} shows the distribution of SI-SDR improvements for the last segment of the test examples when the model operates in \ac{AoTSE} condition for different training strategies and configurations. The model trained with the standard training strategy performed poorly in general compared to the other strategies and failed completely in the causal configuration when using the \ac{cLN}. The model trained with the \ac{MDT} strategy is found to be robust to the choice of the normalization layer and yields better performance compared to the other strategies in both causal and non-causal configurations.
%The performance of \ac{MTT} strategy is comparable to the \ac{MDT} strategy, but fails in the non-causal configuration with \ac{gLN}. However,
%The model trained with the dropout training strategy is robust for all the configurations and yields better performance than standard and \ac{MTT} training strategies. While the model with standard training strategy completely fails for causal configuration with \ac{cLN} normalization, the model with \ac{MTT} training strategy fails for non-causal configuration with \ac{gLN} normalization. 

%\begin{figure}[t]
%    \centering
%\end{figure}

%\begin{figure}[t]
%    \centering
%    \includegraphics[width=0.9\linewidth]{figures/violin_plot_simulated_section_2.pdf}
%    \caption{Violin plots showing the distribution of SI-SDR improvements (SI-SDRi) for the 3rd segment of the test examples for different configurations and training strategies. Mean values are shown on top of the distribution.}
%    \label{fig:distributions}
    %\vspace{-1.0em}
%\end{figure}

\vspace{-4mm}
\section{Conclusion}\label{sec:conclusion}
%\vspace{-3mm}
%In this paper, 
We compared a \ac{MDT} strategy for modality dropout resilient \ac{MTSE} system training to the existing strategies in non-causal and causal configurations. A comparison of these strategies, in combination with the choice of normalization layers used by the DNN, showed that the \ac{MDT} strategy is effective in diverse practical scenarios while the existing strategies are sensitive to the architecture design. %, particularly in the non-causal configuration. 
In addition, we showed that the system trained with the \ac{MDT} strategy can effectively use the extracted speech as an enrolment signal, outperforming other training strategies. Although we investigated clean speech mixtures in the work, we hypothesize that the conclusions hold even for noisy speech mixtures since the model robustly extracts features relevant for target speaker extraction from the auxiliary information streams. Investigating the different fusion strategies and architectural choices is a topic for future research.  

% \section{Acknowledgement}
% % \vspace{-3mm}
% The authors thank the Erlangen Regional Computing Center (RRZE) for providing computing resources and support.

% References
% \newpage

\balance
\bibliographystyle{IEEEtran}
\bibliography{sapref}

\end{document}